\documentclass[journal]{vgtc}
\ifpdf
  \pdfoutput=1\relax                   
  \usepackage{graphicx}                
  \DeclareGraphicsExtensions{.pdf,.png,.jpg,.jpeg} 
\else
  \usepackage{graphicx}                
  \DeclareGraphicsExtensions{.eps}     
\fi%

\vgtcinsertpkg

\graphicspath{{figures/}{pictures/}{images/}{./}} 
\usepackage[utf8]{inputenc}
\usepackage{microtype}                 
\PassOptionsToPackage{warn}{textcomp}  
\usepackage{textcomp}                  
\usepackage{mathptmx}                  
\usepackage{times}                     
\usepackage{cite}                      
\usepackage{tabu}                      
\usepackage{booktabs}                  
\usepackage{subcaption}
\usepackage{multirow}

\onlineid{1080}

\vgtccategory{Research}
\vgtcpapertype{application/design study}

\usepackage{xcolor}

\newcommand{\rev}[1]{{\color{black}{#1}}}

    \title{ChemVA: Interactive Visual Analysis of Chemical Compound Similarity in Virtual Screening}

\author{Mar\'{i}a Virginia Sabando\thanks{These authors contributed equally.} , Pavol Ulbrich\footnotemark[1] , Mat\'{i}as Selzer, Jan By\v{s}ka,\\ Jan Mi\v{c}an, Ignacio Ponzoni, Axel J. Soto, Mar\'{i}a Luj\'{a}n Ganuza, Barbora Kozl\'{i}kov\'{a}}
\authorfooter{
\item 
M. V. Sabando, I. Ponzoni, and A. J. Soto are with the Institute for Computer Science and Engineering (UNS–CONICET) and with the Department of Computer Science and Engineering, Universidad Nacional del Sur, Bah\'{i}a Blanca, Argentina. 
E-mails: virginia.sabando@cs.uns.edu.ar, ip@cs.uns.edu.ar, axel.soto@cs.uns.edu.ar.
\item 
P., Ulbrich, J. By\v{s}ka, and B. Kozl\'{i}kov\'{a} are with the Visitlab, Faculty of Informatics, Masaryk University, Brno, Czech Republic. E-mails: paloulbrich@gmail.com, jan.byska@gmail.com, kozlikova@fi.muni.cz.
\item 
M. N. Selzer and M. L. Ganuza are with the Institute for Computer Science and Engineering (UNS–CONICET) and with the VyGLab Research Laboratory (UNS-CICPBA), Department of Computer Science and Engineering, Universidad Nacional del Sur, Bah\'{i}a Blanca, Argentina.
E-mails: matias.selzer@cs.uns.edu.ar, mlg@cs.uns.edu.ar.
\item 
J. Mi\v{c}an is with the Loschmidt Laboratories, Department of Experimental Biology and RECETOX, and with the Faculty of Medicine, Masaryk University, Brno, Czech Republic. E-mail: honzamicann@gmail.com.
}

\shortauthortitle{Sabando, Ulbrich \MakeLowercase{\textit{et al.}}ChemVA: Interactive VA in Virtual Screening}

\abstract{
In the modern drug discovery process, medicinal chemists deal with the complexity of analysis of large ensembles of candidate molecules.
Computational tools, such as dimensionality reduction (DR) and classification, are commonly used to efficiently process the multidimensional space of features.
These underlying calculations often hinder interpretability of results and prevent experts from assessing the impact of individual molecular features on the resulting representations.
\rev{To provide a solution for scrutinizing such complex data}, we introduce ChemVA, an interactive 
application for the visual exploration of large molecular ensembles and their features.
Our tool consists of multiple coordinated views: Hexagonal view, Detail view, 3D view, Table view, and a newly proposed Difference view designed for the comparison of DR projections.
These views display DR projections combined with biological activity, selected molecular features, and confidence scores for \rev{each of} these projections. \rev{This conjunction of views allows} the user to drill down through the dataset and to efficiently select candidate compounds. 
Our approach was evaluated on two case studies of finding structurally similar ligands with similar binding affinity to a target protein\rev{, as well as on an external qualitative evaluation.} 
The results suggest that our system allows effective visual inspection and comparison of different high-dimensional molecular representations.
Furthermore, ChemVA assist\rev{s} in the identification of candidate compounds while providing information on the certainty behind different molecular representations.}


\CCScatlist{ 
 \CCScat{K.6.1}{Management of Computing and Information Systems}%
{Project and People Management}{Life Cycle};
 \CCScat{K.7.m}{The Computing Profession}{Miscellaneous}{Ethics}
}

\teaser{
 \centering
 \includegraphics[width=410pt]{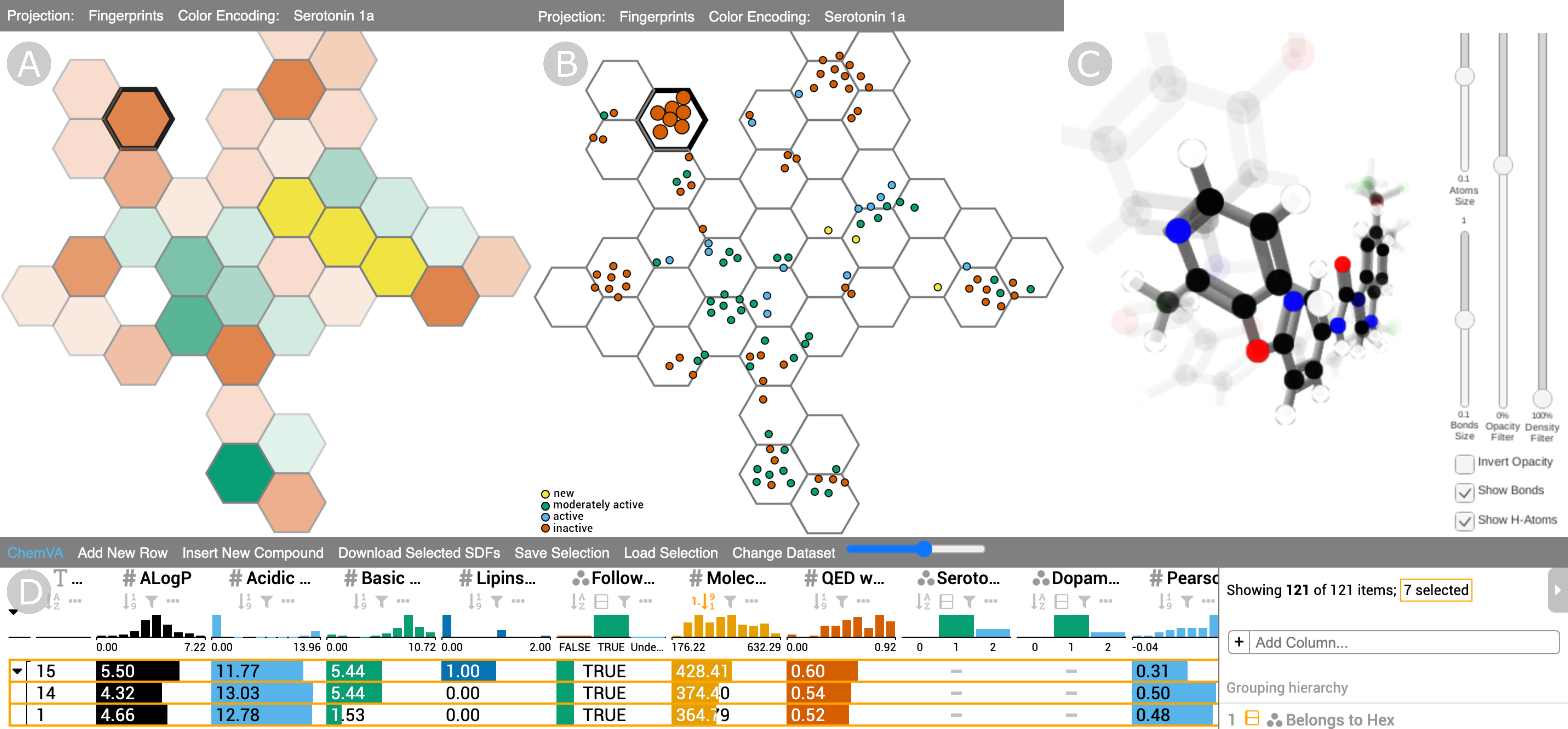}
 \caption{Overview of the ChemVA interface: a) Hexagonal view with an overview of a selected 2D projection, b) Detail view showing all data items, \rev{which are} colored according to a selected feature, c) 3D view enabling the user to observe the structural similarities of selected compounds, d) Table view listing other important features of the compounds. All views are interactively linked.}
 \label{fig:teaser}
 }

\begin{document}



\firstsection{Introduction} 

\maketitle

Small organic chemical compounds are the cornerstone of drug design. New medications are found by exploring a large number of candidate compounds or by  designing new ones. 
In the last decades, high-throughput screening has been the main procedure applied during the early stages of the drug discovery process~\cite{macarron2011impact,hertzberg2000high}. 
    This process requires chemical synthesis, experimental testing 
    of large libraries of \rev{compounds} against a biological target (protein), and it has a high attrition rate, which makes the process costly and time-consuming. 
    These drawbacks stimulated the development of virtual screening methods, which \rev{consist of} computational techniques \rev{for} identify\rev{ing compounds binding} to a drug target. 
    Virtual screening allows to significantly narrow down the number of drug candidate compounds at a faster pace while lowering costs~\cite{Lionta2014,Yu2017}. These reasons make virtual screening an essential part of the early-stage drug discovery process.

    Computational techniques involved in virtual screening enable to simulate and test the fitness of the candidate compound towards the desired function without the need for chemical synthesis and expensive wet laboratory work~\cite{tanrikulu2013holistic}. 
    \rev{They typically }
    consist of a high-dimensional vector-based abstraction of a given molecule, which involves dealing with the challenges inherent in high-dimensional data~\cite{johnstone2009statistical}.
    \rev{Many visualization tools for virtual screening have been proposed in the past years \rev{to help} medicinal chemists deal with the complex computational methods. 
    Such tools have been focused on providing proper means to explore the chemical space and to enhance the interpretability of results so that trusted and explainable decisions can be made~\cite{probst2018exploring,hoffmann2019next}.}
    A common strategy followed by such tools 
    is the application of dimensionality reduction (DR) techniques~\cite{Gutlein2014,probst2018exploring}, which allow mapping compounds from a high-dimensional chemical space into a lower-dimensional (often 2D or 3D) representation~\cite{Gutlein2014}. 
    However, visual exploration of \rev{the} chemical space in the context of virtual screening entails a series of new challenges, not yet addressed by the existing tools. 
    
    \rev{One of these important yet unsolved challenges is to study compound similarity under the premise that similar molecules tend to have similar bioactivity profiles~\cite{johnson1990concepts}.
    Visualization tools should enable the domain expert to interact with different sources of molecular information, \rev{and provide} complementary views that help find similarity determinants. Most existing tools rely on a single mapping based on an arbitrarily chosen set of molecular features.}
    \rev{A visualization tool for virtual screening should also provide views and interactions that allow the domain expert to assess the relationship between the bioactivity profiles of the analyzed compounds and their spatial organization in the projected low-dimensional space.}
    \rev{Lastly, as a consequence of DR, pairwise distances between compounds in the low-dimensional mapping can differ from the corresponding pairwise distances in the high-dimensional space. Information about the trustworthiness of these mappings should be visually presented to the domain expert in an interpretable way. Despite its importance, most of the existing tools do not display this information.}

    These issues were the main driving force behind our research and\rev{,} as a result, we introduce ChemVA, an interactive system for the visual exploration of chemical compounds, targeted for virtual screening. 
    ChemVA provides domain experts with several linked views. 
    One set of views, consisting of 2D plots, supports the visual \rev{inspection} of multiple molecular representations after undergoing DR. 
    Each DR method produces a projection that can be displayed and 
    mutually compared using our newly proposed \textit{Difference view}. 
    Our tool handles the overplotting problem of such projections, which is common for large datasets.
    ChemVA also incorporates a correlation encoding to the plots, which conveys the trustworthiness of a low-dimensional projection based on the distortion with regard to \rev{the pairwise} distances in the original space.
    The plots are complemented by an interactive \textit{Table view}, which allows for sorting and filtering and shows detailed information about the compounds being displayed, focusing primarily on features related to drug-likeness.  
    The tool also contains a \textit{3D view} that allows to explore the structural similarity among selected compounds. 
    This view performs an alignment of selected compounds with respect to their maximum common substructure, which 
    \rev{serves to identify} commonalities and differences among them. 
    Finally, ChemVA allows loading new compounds to an existing dataset, which enables \rev{to} study potential new drug candidates in the context of the dataset under study. 
    The main contributions of this paper can be summarized as follows: 
    \begin{itemize}
        \rev{
        \item A novel visualization tool for virtual screening that allows to compare and contrast multiple dimensionality reduction projections conveying complementary sources of molecular information.
        \item An approach for visually assessing the trustworthiness of each DR projection, which enables the user to focus on the most suitable vector-based molecular representations.
        \item A newly proposed \textit{Difference view} that enables the comparison of multiple DR projections and assessment of their trustworthiness in terms of neighborhood preservation.}
       \rev{
       \item Support for \textit{de novo} drug design by means of a set of coordinated views that allows to analyze newly designed compounds and compare them with an existing set of chemical compounds. 
       \item Functional validation of the proposed tool by means of two case studies and a qualitative testing of the proposed views. These evaluation activities were conducted by three domain experts and one visualization expert.
       }
    \end{itemize}

    


\section{Related Work}
 \label{sec:related_work}
    
    In this \rev{s}ection, we conduct a brief survey \rev{of} existing approaches in the different areas related to our work. 
    \rev{We review DR techniques and their application to the visualization of high-dimensional spaces, focusing mainly on parametric DR methods.}
   Then, we discuss the state-of-the-art molecular visualization and visual exploration tools for large molecular ensembles.

    \subsection{Visual Exploration of Multidimensional Data}

    Over the last few decades, a variety of visualization methods for multidimensional data have been proposed~\cite{liu2016visualizing, borland2019selection,wang2019polarviz}.
    Discovering patterns in multidimensional data using a combination of visual and machine learning techniques represent\rev{s} a well-known challenge in visual analytics. 
    Visual exploration of multidimensional data allows for the injection of unique human perceptual and cognitive abilities directly into the process of discovering multidimensional patterns~\cite{kovalerchuk2018visual}. 
    The challenge mostly lies in having a large number of variables and their relationships that have to be considered simultaneously. 
    As the number of variables increases, the user's ability to understand interactions and correlations between them is severely limited~\cite{wang2019high}.

    In this context, a variety of approaches have been introduced to visually convey high-dimensional data by \rev{using} two-dimensional projections, so that salient structures or patterns can be perceived while exploring the projected data \cite{moon2017visualizing, zanabria2016istar,sacha2016visual}.
    Therefore, one typical approach is to transform the original dataset using a DR technique, and then \rev{to} visually encode only the reduced data~\cite{sedlmair2012dimensionality,8383983}.	
    A frequently used visual encoding technique for showing the projected low dimensional data is a scatter plot.
    Sedlmair et al.~\cite{sedlmair2013empirical} carried out an extensive investigation on the effectiveness of visual encoding choices, including 2D scatter plots, interactive 3D scatter plots, and scatter plot matrices. 
    Their findings suggest that the 2D scatter plot is the most suitable approach to explore the output of different DR algorithms.

    Many DR techniques have been proposed throughout the last decades to transform multidimensional feature spaces onto a low-dimensional manifold---typically onto a two or three-dimensional space---while trying to preserve neighborhood relationships among the original data instances~\cite{van2009dimensionality,xu2019review}.
    Among these methods, \textit{t-distributed Stochastic Neighbor Embedding} (t-SNE)~\cite{maaten2008visualizing} has been widely adopted, specifically for the visualization of very high-dimensional data~\cite{mcinnes2018umap, rauber2016visualizing, tang2016visualizing, wu2017visualization}.
    Many of the proposed DR techniques are non-parametric, i.e.,~they find patterns in specific manifolds and do not provide means to map new data points from the original high-dimensional space to the latent space. 
    Therefore, their out-of-sample extension is not possible. 
    Linear methods for DR, such as PCA~\cite{wold1987principal}, LDA~\cite{fisher1936use}, or MSR~\cite{strickert2010adaptive}, have been extensively used, but are constrained to linear transformations of the original data.  In contrast to other non-linear techniques, e.g.,~SOMs~\cite{kohonen1990self}, MDS~\cite{cox2008multidimensional} or autoencoders~\cite{kramer1991nonlinear}, which allow to map non-linear data, t-SNE tackles the \emph{crowding problem} effectively by using a heavy-tailed distribution for arranging data points in the low dimensional space~\cite{maaten2008visualizing}.
    Practically speaking, addressing the crowding problem yields better-looking visualizations. Although t-SNE is non-parametric, there is a parametric t-SNE version~\cite{van2009learning}, which seeks to overcome this limitation. 
    Nonetheless, it is often difficult to find an optimal configuration of hyperparameters for these models, which in turn yields noisy projections compared to those obtained using the non-parametric version of the algorithm~\cite{min2018parametric, zhu2018generic, xue2018deep}.

    \rev{In particular, v}isualization tools for virtual screening should allow medicinal chemists 
    to analyze how a previously unseen compound interacts with or relates to other known compounds.
    For this reason, DR techniques used in the visualization tools for virtual screening should provide means for mapping unseen compounds and they 
    should be fast enough to enable interactive use.

    \subsection{Visualization of Molecular Structures}
    Molecular visualization is one of the oldest branches of visualization as it has a well-established basis of visual representations, which has been largely embraced by biochemists and biologists. 
    These---mostly 3D---representations are integrated into many available molecular visualization tools, such as PyMOL~\cite{PyMOL}, VMD~\cite{humphrey1996vmd}, Chimera~\cite{chimera}, or YASARA~\cite{krieger2014}. 
    Besides these tools, there are also several others---mainly web-based ones---that can be embedded into other applications; for instance, Jmol~\cite{jmol}, JSmol~\cite{jsmol}, or 3Dmol.js~\cite{3dmol}. However, none of these tools is directly applicable to the problem of visual exploration of large ensembles of molecular structures and the similarities in their structure and properties.
    A more extensive overview of the currently available approaches to molecular visualization and molecular systems can be found in the survey by Kozl\'{i}kov\'{a} et al.~\cite{kozlikova2016star}.
    
    \rev{Visual inspection} of small molecules is supported in different ways by some available tools. Smaller molecules are commonly visualized in 2D in order to better depict their structure, such as in LigPlot+ \cite{ligplot}.
    This tool serves mainly for exploring the interactions between a ligand and its target protein, by projecting its molecular structure and the protein amino acids onto the 2D plane.
    However, such a tool is not applicable to the task of analyzing multiple molecular compounds.

    The exploration of large sets of chemical compounds constitutes an ideal task from the visual analytics viewpoint, and several such tools have already been proposed.
    The closest to our proposed tool is CheS-Mapper \cite{Gutlein2014}.
    It enables to load data from several chemical databases, to calculate molecular descriptors, to observe clusters of compounds in a reduced 2D space, and to display the 3D representation of the compounds.
    Nonetheless, the tool does not provide an option to interactively compare different projections and does not support \rev{inspecting} the physical-chemical properties of a newly added compound. 

    A similar tool is Data Warrior \cite{sander2015datawarrior}, which consists of several basic visualization methods and specialized views to show an overview of the whole chemical and pharmacophore space of the input dataset.
    However, the tool is not designed
    for specific tasks related to the early-stage drug discovery process that is addressed in our tool.
    More recently, Yoshimor\rev{i} et al.~\cite{yoshimori2019integrating} proposed a technique for visually summarizing information about multiple Structure-Activity Relationship Matrices, based on molecular grid maps and 3D activity landscapes. 
    Nevertheless, due to its specificity, it cannot be extended to operate with additional physical-chemical properties and descriptors.

    Janssen et al.~\cite{janssen2018drug} introduced a method that uses a visual mapping based on t-SNE projections for finding new potential kinase inhibitors.
    The views provided by the tool show the results of clustering and a tree-like structure of compounds. 
    However, such views are not designed to easily extract information about the structural similarity of compounds.
    A similar concept 
    was introduced by Probst and Reymond~\cite{probst2019visualization}. 
    Their approach is tailored to process very large high-dimensional datasets and the compound similarity is expressed by the proximity of compounds through tree branches.

    Naveja and Medina-Franco~\cite{naveja2019finding} introduced the constellation graphs of chemical compounds clustered according to a shared core scaffold in the 2D projections of their geometry. 
    Each cluster is annotated with a small view containing the projected core.
    While such constellation graphs provide an innovative approach for the visualization of compounds, they do not fit our requirements for visual exploration in the context of virtual screening.
    Synergy Maps~\cite{lewis2015synergy} is another web-based application for displaying relationships between compounds and understanding potential synergies between them.
    The tool shows pairwise combinations of properties of compounds using a network representation.
    Nonetheless, the node-link diagrams 
    do not scale properly, which hinders its application for 
    large ensembles of compounds.
    
    Although methods for assessing the trustworthiness of DR techniques have been proposed \cite{aupetit2007visualizing, cutura2018viscoder, peltonen2015information}, to the best of our knowledge none of the existing approaches for visualization of molecular structures allows to perform this task. Moreover, they do not provide means to compare the results of multiple projections. 


\section{Background} \label{sec:backgound} 
In this \rev{s}ection, we provide details on the theoretical background of our proposed tool, which includes the molecular representations used and their semantics, as well as a short description of the chemical features of major relevance for drug-likeness.

\subsection{Vector-Based Molecular Representations}
\label{sec:molecular_representations}
Research \rev{in cheminformatics} suggests that there are several different factors involved in assessing chemical compound similarity, which go beyond 
\rev{having similar 
geometric arrangements of atoms and bonds}~\cite{maggiora2004molecular,barbosa2004molecular}. 
Given the relevance of assessing compound similarity in virtual screening, it is important to provide medicinal chemists with a variety of vector-based molecular representations that capture different aspects of the compounds under study and that are complementary to each other. 
In this sense, ChemVA provides four different sources of data: 
\begin{itemize}
    \item \textit{Extended Connectivity Fingerprints} (ECFPs)~\cite{rogers2010extended} encode the topological information of the molecular structure as a fixed-length vector of bits. 
    \item \textit{Daylight Fingerprints} \cite{daylight} are path-based vectors of bits that encode all fragments of a molecule, which are obtained by traversing its molecular graph.
    \item \textit{Molecular Descriptors} \cite{todeschini2008handbook} are numerical values associated with the chemical constitution of a molecule. 
    \rev{They convey information such as molecular weight, electronic configuration, solubility, etc.}
    \item \textit{Molecular Embeddings} \cite{jaeger2018mol2vec} are a rather novel type of vector-based representations. 
    \rev{They} are typically obtained by applying machine learning models that are trained to learn data representations of molecules \rev{from large and diverse sets of compounds}.
\end{itemize}
We selected these four vector-based molecular representations as our sources of data because they capture different information about the compounds, thus 
allowing for a broader analysis of the data under study, and because each of them is considered to be the state-of-the-art of their type. 
While we are utilizing these four vector-based molecular representations, it is worth noting that ChemVA remains independent from the chosen sources of data. Details \rev{of their} calculation 
are provided in the Supplementary Material.

\subsection{Molecular Features Related to Drug-Likeness}
\label{sec:druglikeness}
Some molecular features are of major interest to drug designers and medicinal chemists in the context of virtual screening.
In particular, molecular features related to drug-likeness allow scrutinizing the compounds in terms of their viability as potential new drugs.
ChemVA takes into consideration a subset of such features:
\vspace{-1mm}
\begin{itemize}
    \item \textit{Molecular weight} represents the average mass of the molecule expressed in Daltons (Da). Drug-like compounds are expected to weigh from $160$ to $500$ Da.\vspace{-0.5mm}
    \item \textit{logP} is a quantitative measure of lipophilicity, 
    \rev{and indicates how easily a substance is absorbed by living tissue.}
    Drug-like compounds \rev{often} exhibit logP values from $-0.4$ to $+5.6$.\vspace{-0.5mm}
    
    \item \textit{Acidic Dissociation Constant} and \textit{Basic Dissociation Constant} are quantitative measures of the strength of an acid (or base) in a chemical solution. 
    \rev{They} are good indicators for a drug\rev{'s} 
    ability to enter the bloodstream 
    \rev{and to accumulate in tissues or secretions.}
    
    \item \textit{Lipinski’s Rule of 5 (RO5)} is a rule of thumb to determine whether a compound 
    \rev{is likely to be active in humans.} 
    This rule states that an orally active drug violates no more than one of four criteria 
    \rev{based on threshold values of specific chemical features.}
    
    \item \textit{Weighed Quantitative Estimate of Drug-likeness (QED)} is a score computed based on chemical features linked to drug-likeness,  
    \rev{and it ranges from 0 (\textit{non drug-like}) to 1 (\textit{drug-like}).}
    
\end{itemize}
\vspace{-1mm}
\section{Requirements}
\label{sec:requirements}

    
    
   
   \rev{Over the course of one year, we conducted numerous sessions with the group of protein engineers from the Loschmidt Laboratories at Masaryk University. Based on their input, we identified several limitations of the existing approaches for virtual screening and summarized them into a list of requirements. The experts agreed that these requirements cover the most critical aspects of a virtual screening workflow. One of the experts, who is also the co-author of this paper, has dedicated significant time ensuring that our implementation addresses the requirements by iteratively checking and commenting on the progress. His 4-year research experience in protein engineering, along with his internal consultations with the head of his group, provided the necessary domain knowledge for the design of the tool.
   }
    
    \vspace{1mm}
    \noindent\textbf{R1: Overview and detailed \rev{analysis} of \rev{a} molecular ensemble in the low-dimensional space.}
    For large datasets, scatter plots, which are commonly used to represent the DR output, may suffer from occlusion problems \rev{for large datasets}. 
    Therefore, the tool should provide visual support for the \rev{analysis} of data on different levels of abstraction, from the overall distribution of the compounds within the 2D space to the detailed view of individual compounds for a selected region of interest.
    
    \vspace{1mm}
    \noindent\textbf{R2: \rev{Visual inspection} of multiple projections.} 
    A set of compounds can be expressed by different vector-based molecular representations, each \rev{yielding} a different DR projection.
    The tool should enable the user to intuitively combine information encoded in the individual projections to allow studying at once the similarity between compounds based on different molecular representations.
    More specifically, this includes exploring similarities and differences between chemical compounds, expressed by the different projections.
    Additionally, the visual representations and interactions should help the domain experts evaluate the suitability of the \rev{selected} DR model.

    \vspace{1mm}
    \noindent\textbf{R3: \rev{Evaluation} of the trustworthiness of projections.}
    \rev{Users need} proper visual support for \rev{assessing} the trustworthiness of a low-dimensional projection based on the distortion with regard to pairwise distances between compounds in the original space. 
    When such trustworthiness can be compared on different DR projections, 
    \rev{users} can focus the exploration on a subset of molecular representations. 
   
    \vspace{1mm}
    \noindent\textbf{R\rev{4}: \rev{Comparison of compounds according to} features related to drug-likeness.}
    Chemical compounds have many features and descriptors related to drug-likeness 
    that can complement other molecular representations in the task of virtual screening.
    Therefore, it is desirable to provide users with an option to \rev{visualize} these additional features along with the projections.
    
    \vspace{1mm}
    
    \noindent\textbf{R\rev{5}: \rev{Comprehensible viewing} of 3D structural similarity.}
    The tool should support the inspection of individual compounds in terms of their 3D geometry, as well as the visual comparison of common 3D substructures in a selected set of compounds.
    For multiple compounds, such a view should convey the information about similarities and differences in their 3D structure.
   
    \vspace{1mm}
    
    \noindent\textbf{R6: \rev{Possibility to add new compounds and comparison with the existing data.}}
    The tool should support the process of exploration of different features and bioactivity \rev{for newly added compounds}.
    The new compound should be projected using the DR model and integrated into the remaining views, so that the user can compare its features to those of the compounds in the \rev{existing} dataset.


\section{ChemVA Design and Implementation}
\label{sec:design_and_implementation}


The design process of ChemVA followed the requirements listed in Section~\ref{sec:requirements}. 
These requirements are addressed through several coordinated views, which pose challenges to the layout of the tool.
ChemVA was developed in JavaScript using D3.js v5 \cite{bostock2011d3} and a Node.js server \cite{tilkov2010node} built using the Express framework and REST API for the connection to supporting services. We used Unity3d \cite{unity} ported to WebGL \cite{webglGl} for the development of 3D components. Two back-end web services, used for computing the alignment of 3D structures and for computing the 2D coordinates of newly added compounds, were developed using Flask web framework \cite{flask} and written in Python v3. Excluding one of the views~\cite{Furmanova_taggle}, all of the functionality was designed and written in-house, prioritizing the system responsiveness and real-time experience with interactions.

In this \rev{s}ection, we first describe the individual views supported by ChemVA in detail, then we briefly introduce the DR model applied in our tool, and lastly we present the main functionality supporting the analysis of newly added compounds.

\subsection{ChemVA Views}
The initial layout, depicted in Figure~\ref{fig:teaser}, contains \rev{the }Hexagonal view, Detail view, and 3D view\rev{ (Figure~\ref{fig:teaser} A, B, and C, respectively}) in the top row of the canvas, whereas the bottom row contains the Table view\rev{ (Figure~\ref{fig:teaser} D)}. \rev{These views account for requirements R1, R4 and R5.}
For tasks involving the comparison of projections, stated in requirement R2, 
additional views need to be displayed. 
After careful consideration, we decided to position these views between the two rows of the basic layout (see Figure \ref{fig:new_compounds}), as they need to be as close as possible to the default 2D plot views.
This row can be shown or collapsed on demand and allows the user to choose which views should be included (e.g., Hexagonal, Detail, or Difference views).
The purpose of this layout is to allow the user to observe the data from different perspectives. In particular, the Difference view shows the differences between the compound neighborhoods from one projection into another, thus allowing the user to compare different projections.
\rev{The tool uses the color scheme proposed by Okabe \& Ito \cite{ichihara2008color}, so it is accessible to users with color vision deficiency.}

\subsubsection{2D Plot Views}
\label{sec:2d_plots}
The core of ChemVA consists of 2D plots, which give the user an overview of the distribution of compounds in a selected DR projection using a given molecular representation. 
This overview is supported by the \textit{Hexagonal view}, a well-adopted and commonly used approach to visualize the outcome of DR techniques~\cite{sedlmair2013empirical}.
The Hexagonal view aims to overcome 
the overplotting problem, in which the projected data items overlap and cause visual clutter \rev{, thus limiting} the interpretability, especially for datasets evincing high similarity between data items.
The Hexagonal view of ChemVA seeks to solve this problem by aggregating individual data items into individual hexagons.
The user can interactively select a subset of hexagons of interest and explore the distribution of individual data items within the \textit{Detail view}. \rev{The combination of the \textit{Hexagonal view} and the \textit{Detail view} aims to fulfill requirement R1.}
Finally, since ChemVA is tailored to support the visual comparison of different projections, it also offers the \textit{Difference view}. This view was specifically designed to address \rev{that} task, \rev{which} is stated in requirement\rev{s} R2 \rev{and R3}.\\

\noindent\textbf{Hexagonal View}\\
    
\begin{figure}[tb]
\includegraphics[width=1.0\linewidth]{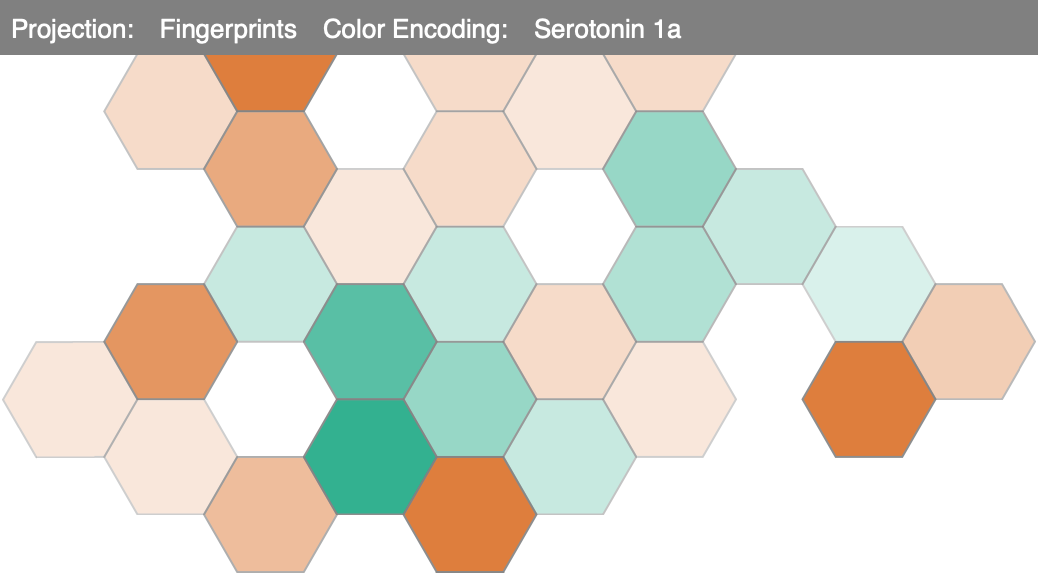}
\caption{\rev{The Hexagonal view shows} the density of the distribution of compounds in the 2D projection, \rev{which is encoded using} into opacity, and the prevalence of the bioactivity of the compounds, \rev{which is encoded using} color. The size of the hexagons can be adjusted.}
\label{fig:hexagonal}
\end{figure}

The Hexagonal view (Figure~\ref{fig:hexagonal}) provides the users with an overview of one particular 2D projection of the dataset. In order to avoid overplotting when dealing with large datasets, the chemical compounds are aggregated into hexagonal bins.
We opted for the hexagonal binning approach, first described by Carr et al.~\cite{carr1987scatterplot}, as it evidences significant advantages for efficient data aggregation in comparison to other approaches. This is related to the low perimeter-to-area ratio of the hexagonal shape, which reduces the sampling bias. The ideal shape for that is a circle but it cannot be used for the full-coverage division of the plane. Hexagons are the most circular-shaped objects, enabling to construct such a division in an efficient way.

The Hexagonal view can be applied to any of the vector-based molecular representations described in Section \ref{sec:molecular_representations}. 
Each hexagon has an assigned opacity, which corresponds to the number of compounds aggregated within the hexagon. A higher opacity corresponds to a larger number of compounds inside the hexagon. In this way, the Hexagonal view depicts the distribution of the compounds yielded by the DR, so \rev{users} \rev{can} recognize areas with a high density of compounds. The color of a hexagon encodes the prevailing trend among its compounds for a selected feature, which is by default their bioactivity \rev{but can be switched to other molecular properties, including the trustworthiness of the projection (requirement R3).}

The granularity of the Hexagonal view can be changed using a slider, which enables a more detailed view of the distribution of the compounds. In order to preserve the readability through the range of different granularity levels, the opacity is enhanced linearly with respect to the decrease of the hexagon size. 
Additionally, the Hexagonal view 
enables the user to select several hexagons at once. The compounds within the selected hexagons are then filtered \rev{through} the other linked views (Detail view, 3D view, and Table view), where they can be explored in more detail.\\


\noindent\textbf{Detail View}\\

\begin{figure}[b!]
    \includegraphics[width=1.0\linewidth]{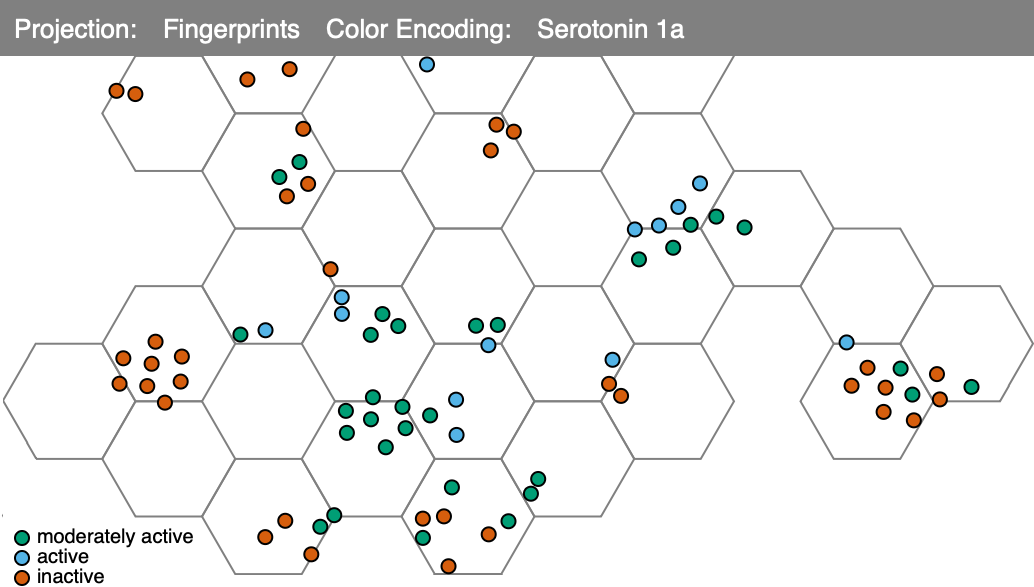}
    \caption{\rev{The} Detail view displays individual compounds as dots in a selected projection (ECFP \rev{fingerprints} in this case). The color can be mapped into one of the features (activity towards Serotonin 1a  receptor in this case).}
    \label{fig:detail}
\end{figure}

Upon selecting a subset of \rev{compounds} in the Hexagonal view, the user can explore the selected \rev{data} in the Detail view, depicted in Figure~\ref{fig:detail}. In this view, the compounds are represented using a standard scatter plot, enhanced by a subtle overlay of the same hexagonal grid as in the Hexagonal view, which helps users keep the correspondence between the zoom level in these views. This is important because the Detail view displays only the selected hexagons zoomed in after the selection.
To further enhance the link between the Hexagonal and Detail views, the corresponding hexagons are highlighted when the user hovers over them in any of these views.
In order to perform a selection of individual compounds in this view, a lasso-shaped selector is supported. The selected compounds are then displayed in the 3D view and also highlighted in the Table view \rev{(Section \ref{sec:list_view})}.

    
As in the case of the Hexagonal view, the Detail view can be set to any of the vector-based molecular representations described in Section \ref{sec:molecular_representations}. In addition, all properties and features described in Section \ref{sec:backgound} can be color-encoded on points representing each compound.
These features are selected from a drop-down menu, and their color encodings are chosen according to their type, i.e., quantitative, such as molecular weight, or categorical, such as bioactivity towards a target protein.
Another quantitative property \rev{that can be used for color encoding} corresponds to the correlation scores, which encode the trustworthiness of the DR projection of the compound \rev{(requirement R3)}. 
These scores were computed using Pearson and Kendall correlation, whose calculation is explained in the Supplementary Material. \\


    
\noindent\textbf{Difference View}\\

In order to support the task of comparing the outputs of different 2D projections, as stated in our requirement R2, we propose a novel view called \textit{Difference view}, 
that combines and contrasts two selected 2D Hexagonal views, $A$ and $B$.
Initially it displays a hexagonal layout similar to that presented in the Hexagonal view, where the opacity of each hexagon encodes the computed correlation score of the trustworthiness of the projections $A$ and $B$ under study \rev{(requirement R3).}

\begin{figure}[t!]
    \includegraphics[width=1.0\linewidth]{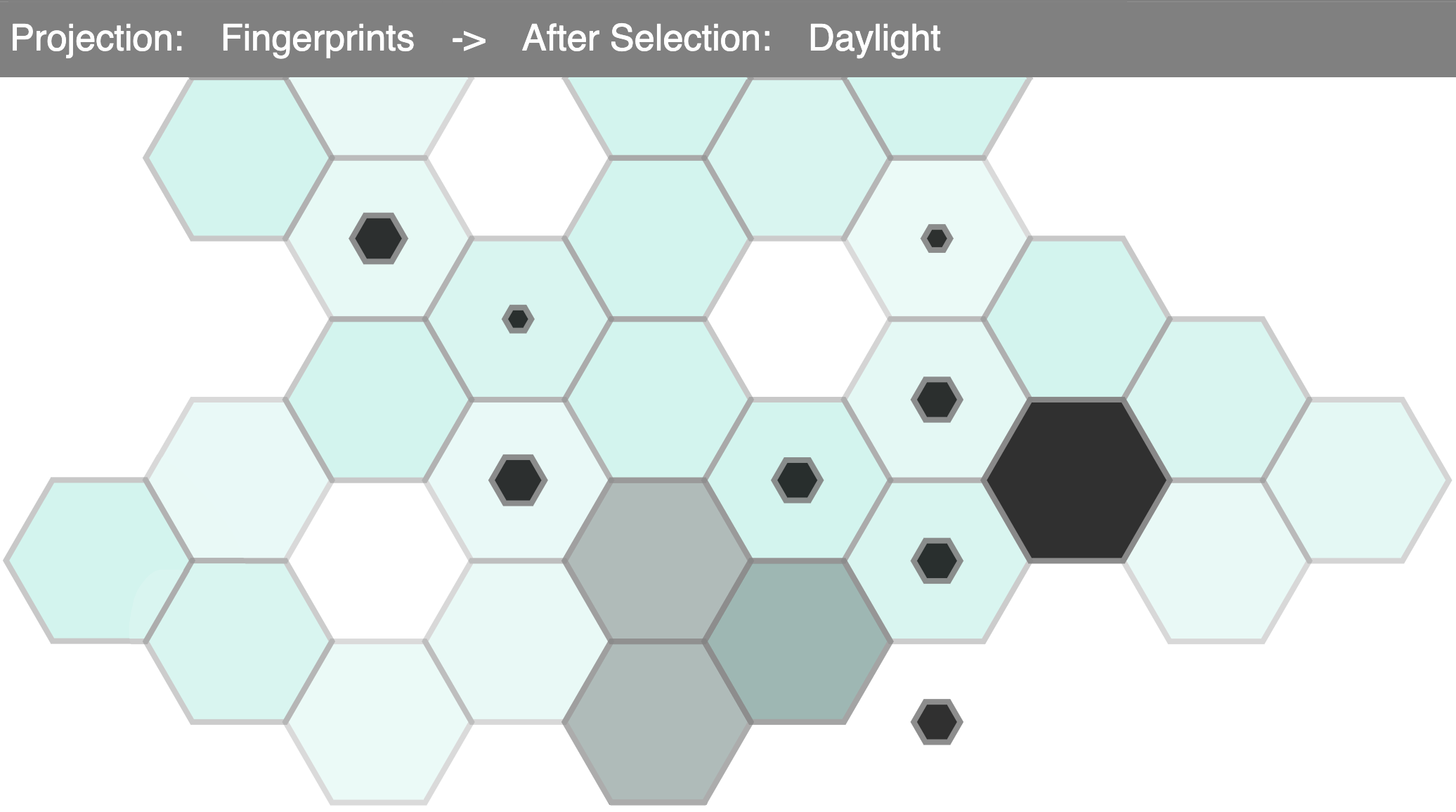}
    \caption{The Difference view showing the decomposition of the local neighborhoods from one projection to another one. Here, the reference projection $A$ corresponds to ECFP fingerprints and its selected hexagons are highlighted in gray. Projection $B$ corresponds to Daylight fingerprints (inner hexagons in black). The opacity of hexagons in $A$ encodes the value of the metric used for the trustworthiness of projection (see Section \ref{sec:case_study}).}
    \label{fig:difference}
\end{figure}

The Difference view adopts the hexagonal layout from one of the projections to be compared, which we denote as the reference projection. When we choose projection $A$ as the reference and perform a selection operation in the Hexagonal view of $A$, we search for the positions of the compounds falling into this selection in projection $B$. These positions are encoded as inner smaller hexagons inscribed into the original hexagonal grid of the Difference view, as depicted in Figure~\ref{fig:difference}. In other words, this view shows where the compounds from the selected hexagons in projection $A$ are located in projection $B$.
The size of inner hexagons corresponds to the number of compounds falling into the same hexagonal bin.


The primary purpose of this graph is to help illustrate which regions of the dataset preserve their neighborhoods from one vector-based molecular representation to another one when undergoing a DR procedure.
In this way, the user can rapidly compare two different projections and assess the trustworthiness of the neighborhoods generated by the DR techniques.
If the level of fragmentation from one projection to another is high, i.e., a segregation of the content of one hexagon into many small, fairly separated and scattered hexagons is observed, it can be inferred that these molecules behave differently according to the chosen molecular representation, thus they might be taken into consideration cautiously and inspected in more detail.


\subsubsection{Table View}
    \label{sec:list_view}
Besides from the vector-based molecular representations used in the DR projections and displayed in our 2D plot views, there are several other molecular features related to drug-likeness that \rev{should be} taken into consideration when \rev{analyzing} the compounds, as stated in requirement~\rev{R4}. ChemVA enables the users to explore \rev{such} features, listed in Section~\ref{sec:druglikeness}, by \rev{means of} a Table view which offers advanced interaction options. 
We adopted a well-established tool, published by Gratzl et al.~\cite{Gratzl_lineup_old}, and its extension~\cite{Furmanova_taggle}.
As these tools perfectly fit to our needs, we incorporate them to ChemVA. Further details about the broad range of interaction possibilities can be found in the original papers. 

\begin{figure}[t!]
    \includegraphics[width=1.0\linewidth]{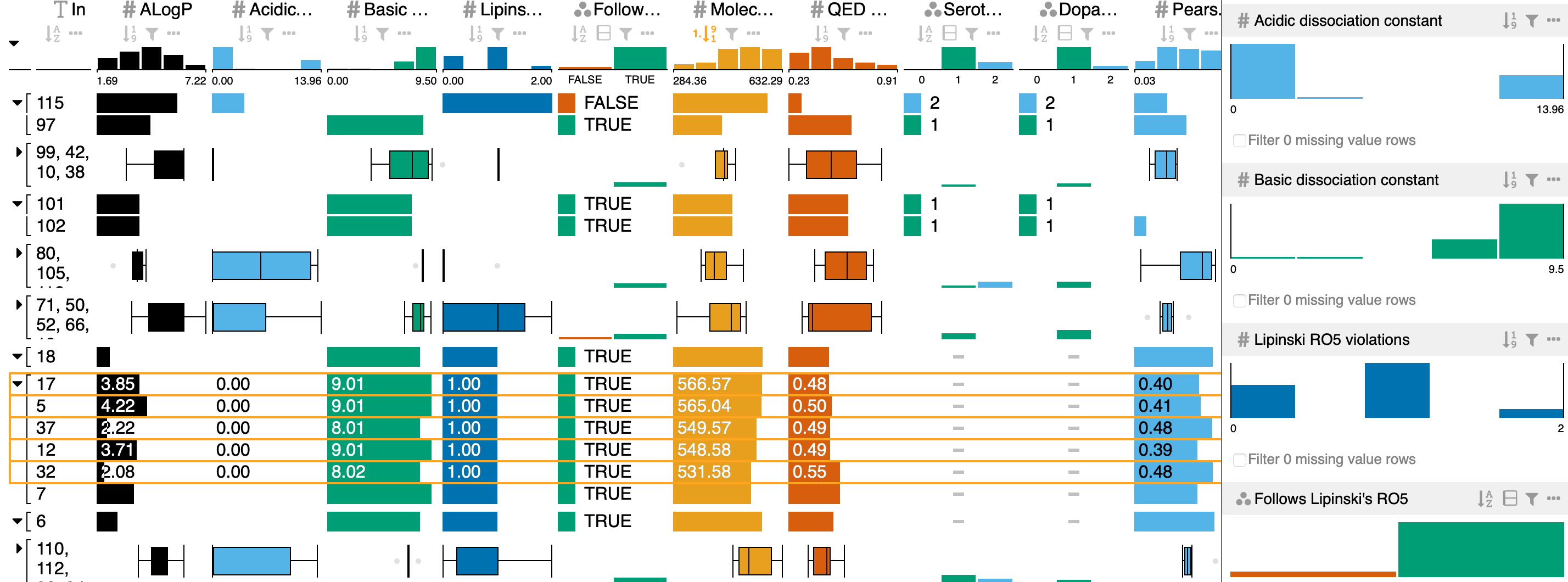}
    \caption{The Table view. Each row corresponds to one compound and its selected features. The right side panel shows the statistical overview of the distribution of values across the dataset.}
    \label{fig:interactive_table}
\end{figure}

In addition to the list of compounds, the Table view provides the users with graphical elements in the form of juxtaposed bar charts and box plots in the right side panel. 
By default, the compounds in the table are logically grouped by their membership to hexagons in the Hexagonal view. These groups can be either expanded or compressed. \rev{When} compressed, the table displays the box plots of the distribution of the values for each feature in the hexagon, as shown in Figure~\ref{fig:interactive_table}.

This view is interactively linked with the other visual components of ChemVA. When performing a selection in the 2D plot views, the corresponding compounds are automatically highlighted in the Table view. Conversely, when the user selects compounds in the Table view, the corresponding compounds are highlighted in the Detail view, displayed in the 3D view, and the corresponding hexagons are highlighted in the Hexagonal view.



    
    


    

\subsubsection{3D View} 
\label{sec:3dview}
\rev{The} geometric similarity between \rev{selected} compounds  
can be explored using our 3D view. \rev{T}he visual representation of atoms and bonds, as well as the coloring \rev{of} the compounds, \rev{are} based on standard representations used in molecular chemistry, with \rev{specific} colors reserved for \rev{each} chemical element.  
\rev{It} also offers \rev{standard} interactions, such as panning, zooming, and \rev{rotation of} the displayed structures.

\begin{figure}[b!]
\includegraphics[width=1.0\linewidth]{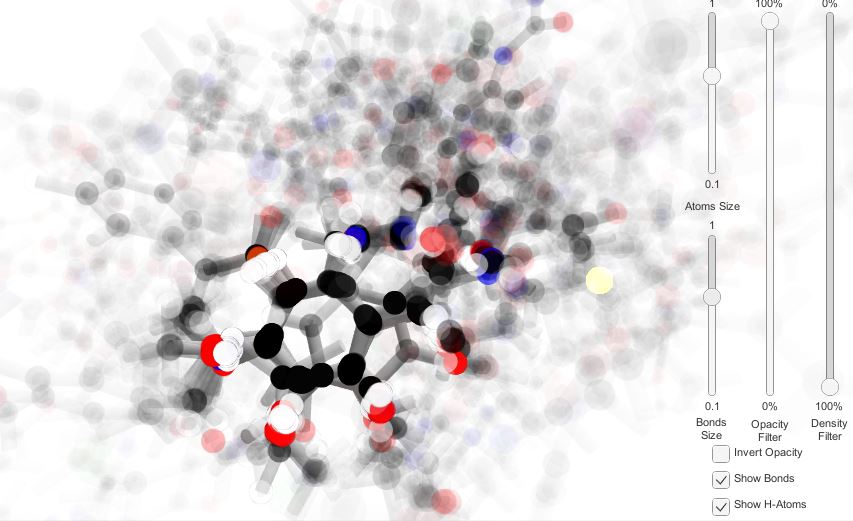}
\caption{3D view with 66 aligned molecules. Common atoms and bonds are rendered more opaque than the less common components.}
\label{fig:3dview}
\end{figure}

Compound similarity can be better perceived when the compounds are structurally aligned in the view.
To serve this purpose, we use a structural alignment functionality provided by the OpenBabel tool~\cite{OBoyle2011}. Further details about this functionality are provided in the Supplementary Material.
Once molecules are aligned, the user should be able to easily identify their common parts, i.e., the subsets of atoms and bonds that are present in most of the selected compounds, as stated in requirement \rev{R5}. 
In addition, we incorporate opacity modulation with respect to the frequency of occurrence of atoms and bonds.
In other words, the opacity of atoms and bonds is calculated based on the number of atoms of the same type that are aligned to the same spatial position. 
As a consequence, common substructures are rendered more opaque. An example of such alignment can be seen in Figure~\ref{fig:3dview}. 

\begin{figure}[t]
    \begin{subfigure}{.25\textwidth}
      \centering
      \includegraphics[width=1.0\linewidth]{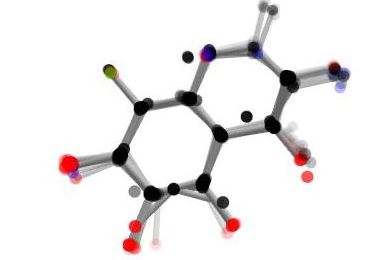}
      \caption{}
    \end{subfigure}
    \begin{subfigure}{.25\textwidth}
      \centering
      \includegraphics[width=1.0\linewidth]{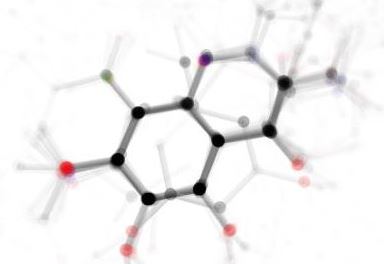}
       \caption{}
    \end{subfigure}
    \caption{Reducing the visual clutter visible in Figure~\ref{fig:3dview} by (a) filtering out the non-common parts of compounds and by (b) changing the opacity of the whole structure.}
    \label{fig:3dFiltering}
    \end{figure}

Depending on the number of selected compounds and their similarities, the 3D visual representation can become fairly cluttered (Figure~\ref{fig:3dview}). 
To reduce this visual clutter, the 3D view offers hiding and showing hydrogen atoms and bonds on demand, changing the size of atoms and bonds, as it can be seen in Figure~\ref{fig:3dFiltering}(a), and also changing the opacity of the whole structure.
Figure~\ref{fig:3dFiltering}(b) shows an example of using these functionalities on a cluttered subset of compounds. 
    
\rev{In some cases,} the user wants to \rev{focus} only \rev{on} the common parts of the structure\rev{.} 
\rev{Therefore}, we have also included functionality for filtering out atoms and bonds. Figure~\ref{fig:3dFiltering}(a) shows how the visibility of the common substructure among 66 compounds is improved by using this feature.
Finally, as the user might also want to analyze the non-common parts of the compounds, the 3D view provides \rev{an option} to invert the opacity, so that the common part of the structure becomes more transparent than the rest.

\subsection{Dimensionality Reduction}
\label{sec:dimensionality_reduction}


\rev{\textit{t-Distributed Stochastic Neighborhood Embedding} (t-SNE) \cite{maaten2008visualizing} is considered to be the state-of-the-art technique for dimensionality reduction, particularly for visualizing very high dimensional data. t-SNE maps neighboring data points to a lower-dimensional space, aiming to preserve the neighborhood relationships (\textit{locality)}. It applies the Student's t-distribution on the low-dimensional space aiming to tackle the \textit{crowding problem}, which happens when many points clump together in the low-dimensional projection. These traits make t-SNE a suitable tool for visually exploring molecular spaces, where local neighborhoods are of utmost interest for analyzing compound similarity.

However, t-SNE has a major limitation, which is that it is a non-parametric technique. After testing \textit{Parametric t-SNE} \cite{van2009learning} and consistently failing to attain good quality projections, we developed our own parametric dimensionality reduction model. This model was constructed based on a feed-forward neural network, which was trained to learn the 2-dimensional coordinates of a previously computed t-SNE projection of the data. We trained four of these parametric models for each reference dataset used in ChemVA, one for each molecular representation. The performance of our parametric models was evaluated by measuring the Pearson correlation coefficient between the predicted coordinates and the actual coordinates, i.e., those obtained by means of the t-SNE projection. 
We provide details on the parameterization used for t-SNE and the parametric models in the Supplementary Material. The t-SNE projections were computed using tools from Scikit-Learn \cite{scikit-learn}, whereas the parametric models were built using Keras and Tensorflow \cite{chollet2015keras}.
}

\subsection{Adding New Compounds}
 \label{sec:new_compound}
ChemVA provides an option to add new compounds to the dataset being studied in order to explore \rev{their} features and compare them with \rev{those of} other compounds. By means of this functionality, the expert can assess the potential of this newly added compound prior to extensive wet lab testing. The support for this important feature fulfills requirement R6. 

A new compound is loaded to ChemVA by specifying its \rev{SMILES formula~\cite{smiles}}. A back-end service of ChemVA calculates the 
molecular features listed in Section~\ref{sec:druglikeness}. 
After computing such molecular representations and features, \rev{ChemVA uses} the parametric models \rev{(described in Section~\ref{sec:dimensionality_reduction})} to obtain the 2D coordinates of the compound in each projection.
Finally, the newly added compound is displayed in all the supported views with yellow color to ease its identification. This color encoding prevails throughout the views\rev{, as it can be seen in Figure \ref{fig:new_compounds}.} 
\section{Evaluation and Case \rev{Studies}}
\label{sec:case_study}

\rev{The evaluation of ChemVA consisted of two main stages. The first stage was conducted by one domain expert involved in the functional design of the tool (see Section \ref{sec:requirements}), and consisted of two case studies. 
The second stage was conducted by one visualization expert and two domain experts, who were not involved at any point during the design or development of ChemVA. This stage consisted of a 
session where the different views where qualitatively evaluated and general feedback about the usability and functionality of the tool was gathered afterwards.}

\rev{\subsection{First Stage - Case Studies}\label{sec:first_stage}}

During the \rev{first} stage of the evaluation, 
\rev{several additional functional requirements for the tool itself were identified by the domain expert, such as loading and storing selections, downloading the displayed 3D conformations, and highlighting the compounds in the Table view after clicking on them in the 3D view. These additional functional requirements were addressed in time for the case study evaluation.}

ChemVA was tested on two case studies, which were built upon two different datasets retrieved from ChEMBL \cite{gaulton2017chembl}.
The first dataset was composed by merging ligands binding to the \textit{Serotonin 1a receptor}\footnote{https://www.ebi.ac.uk/chembl/target\_report\_card/CHEMBL214/} and \textit{Dopamine D2 receptor}\footnote{https://www.ebi.ac.uk/chembl/target\_report\_card/CHEMBL217/}, whereas the second dataset comprised ligands to the \textit{P-glycoprotein~1}\footnote{https://www.ebi.ac.uk/chembl/target\_report\_card/CHEMBL4302/}. 
We assigned a categorical label to each compound according to its experimentally measured \textit{IC50} bioactivity value towards the target(s) under study. Compounds showing \textit{IC50} values \rev{below $10\ nM$ were labeled as \textit{Active}; compounds between $10$ and $1000\ nM$ were labeled as \textit{Moderately Active}, and those over $1000\ nM$ were labeled as \textit{Inactive}.}
The \textit{serotonin-dopamine} dataset comprise\rev{s} 118 compounds, whereas the \textit{P-glycoprotein} dataset contain\rev{s} 893 compounds.

\subsubsection{Case Study 1: \rev{Analysis of Chemical Determinants of Activity Towards Serotonin and Dopamine Receptors}}

\begin{figure*}[ht]
    \centering
    \includegraphics[width=\textwidth]{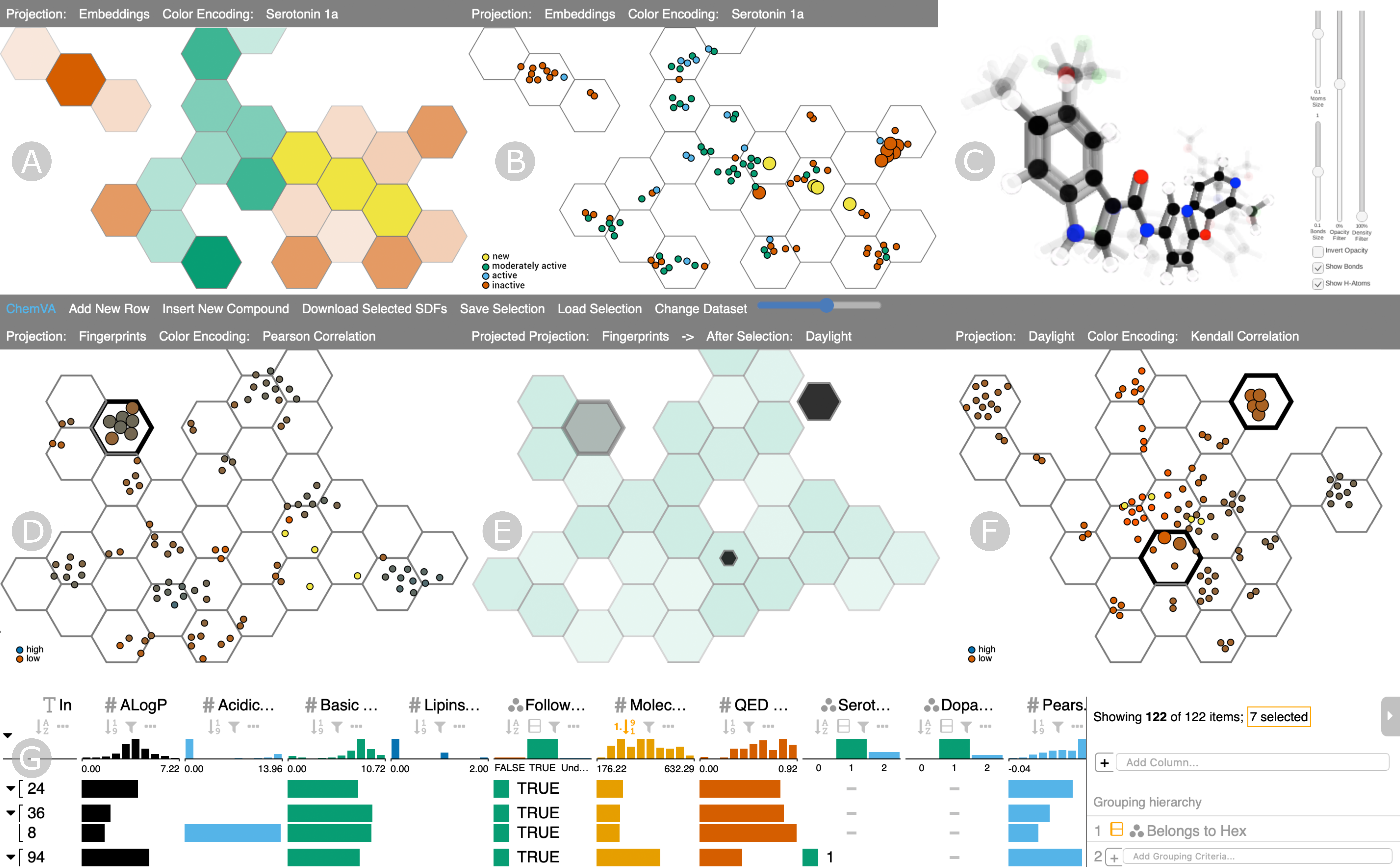}
    \caption{The newly designed compounds were projected by ChemVA near the selected dataset compounds in the Daylight projection (top), occupying the same hexagons, and slightly further from them in the ECFP Fingerprints projection (bottom). The Difference view on the bottom right illustrates the redistribution of the selected regions. a) Hexagonal view, b-d-f) Detail view, c) 3D view, e) Difference view, g) Table view. }
    \label{fig:new_compounds}
    \vspace{-5mm}
\end{figure*}

This case study is based on the \textit{serotonin-dopamine} dataset, a set of compounds that have data about antagonistic activity against the serotonin 5HT1A and dopamine D2 receptor.
These receptors of neurotransmitters are targets of many psychoactive pharmaceuticals, i.e., antidepressants, antipsychotics, and anxiolytics. 
In the field of psychopharmacology, small molecules with different activity towards several receptors are used to alleviate side-effects \cite{nasrallah2008atypical, siafis2018antipsychotic}.

The goal of the study was to find chemical determinants of biological activity towards serotonin and dopamine receptors.
According to the domain expert, similar compounds \rev{could} be easily found in the Detail view in all four projections, based on their proximity.
After identifying groups of potentially similar compounds, the domain expert searched for groups of compounds that were \textit{active} towards both receptors and that had desirable pharmacological properties, such as following the Lipinski‘s RO5 and having a high QED score.
This exploratory search was \rev{done on} the Hexagonal view, by adjusting the size of the hexagons to match the observed groups of compounds in the different projections.

\rev{Afterwards}, the domain expert used the Kendall and Pearson correlation color encoding in order to observe whether the projections were trustworthy, and thus the hexagons being considered would effectively group similar compounds.
A set of hexagons was preselected and the Table view was used, in which the domain expert analyzed the distributions of all drug-likeness features of the selected compounds using the summary information displayed in the column headers. 
\rev{As a result, the expert identified a hexagon that grouped active compounds towards both receptors with desirable drug-likeness features.}

The domain expert selected those compounds and performed an alignment in the 3D view, finding that their structures \rev{were} very similar.
These compounds were also contrasted to inactive compounds within the same hexagon using the 3D view.
This allowed the domain expert to gain insight into the relevant substructures of a potential new drug candidate.
At this point, the domain expert highlighted 
how easy it was to find and visualize the common 3D structure of a group of compounds.   

\rev{Finally,} the structures of active compounds towards both receptors were downloaded from ChemVA. 
Five new compounds were created and loaded to ChemVA, as shown in Figure \ref{fig:new_compounds}.
They were compared to the downloaded structures using molecular docking. This process is described in the Supplementary Material.
\rev{The domain expert found out that the new structures effectively bound to a known antagonist binding pocket and three out of the five newly created structures showed slightly better binding energies than structures from the dataset used for their design, which were already highly active.} This \rev{shows} that ChemVA could be effectively used for drug design purposes, leading to newly designed compounds that have the desired qualities and bioactivity profiles.

\subsubsection{Case Study 2: Analysis of Structural Determinants of P-glycoprotein Inhibitors}

\begin{figure*}[ht]
    \centering
    \includegraphics[width=\textwidth]{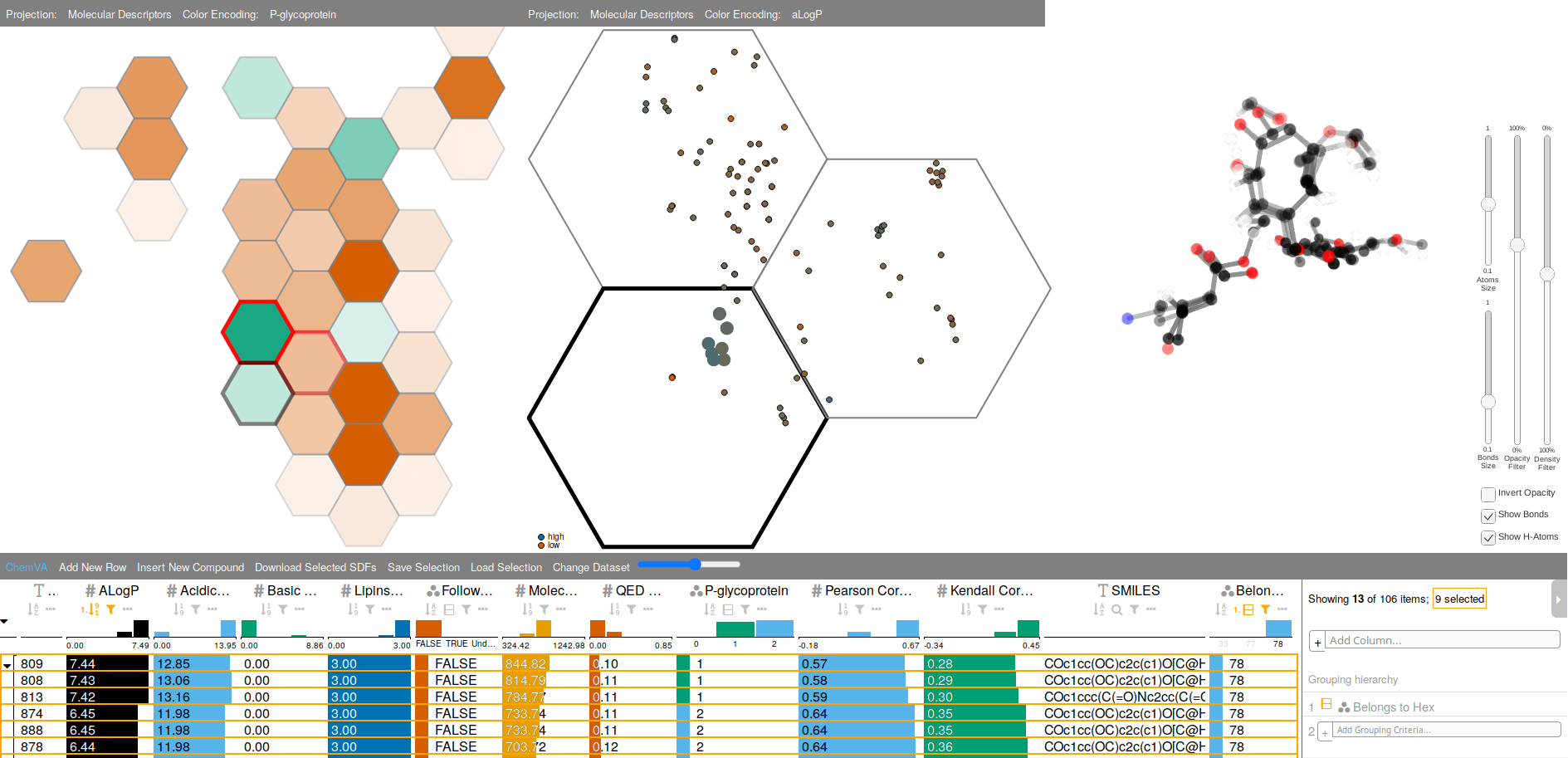}
    \caption{Illustration of the exploratory process in the P-glycoprotein dataset, displaying only a subset of interesting molecules in the Detail view and selecting \rev{those} with high logP values.}
    \label{fig:2nd_case}
    \vspace{-5mm}
\end{figure*}

This case study was based on the \textit{P-glycoprotein} dataset, consisting of small molecules with inhibitory activity against human P-glycoprotein. 
\rev{This protein is} exclusively overexpressed in cells of many cancer types,  
\rev{causing} multidrug resistance of these cells \rev{and thus impacting on the performance of chemotherapy treatments}. \rev{It} also affects the effectivity of many drugs by altering their \rev{ADME-Tox properties.}
\cite{srivalli2012overview}. 
\rev{P-glycoprotein} has proven to bind to many structurally dissimilar substrates, thus an interesting use case for our tool was to find structural determinants of a good P-glycoprotein inhibitor and to compare known active ligands towards this target.

According to the domain expert, the P-glycoprotein \rev{dataset} contains a vast number of diverse compounds that interact with the target protein.
The goal of this case study was to find chemical determinants of compounds with a very high logP value, which are scarce in the dataset.
In order to achieve this goal, the domain expert used the Table view \rev{to filter} compounds with logP values higher than 6.75 and then sorted them based on their logP value and their correlation scores, which \rev{measure} the trustworthiness of the molecular representation currently being projected onto the 2D plot views.

The domain expert used the filtering options provided by the Table view in order to find a set of compounds complying with all the requested criteria, i.e., high logP value, activity towards the P-glycoprotein and following Lipinski‘s RO5.
Afterwards, chemically similar compounds were selected from this subset and further analyzed using the Hexagonal view. Hexagons containing active compounds were selected and thoroughly explored using the Detail view, in which logP was used for color encoding. Finally, the domain expert used the 3D view to find common substructures among the selected compounds and successfully identif\rev{ied} chemical determinants of bioactivity. This was achieved by analyzing structural patterns related to lipophilicity, indicated by high logP values. This workflow is depicted in Figure \ref{fig:2nd_case}. 

\rev{\subsection{Second Stage - Qualitative Evaluation by External Experts} \label{sec:second_stage}

In the second stage we introduced ChemVA to two domain experts from the Loschmidt Laboratories at Masaryk University who evaluated our tool in terms of functionality and user-experience. We also introduced our newly proposed Difference graph to one visualization expert (see Acknowledgments), who has vast experience in DR techniques. None of the experts were involved at any stage during the design and development of ChemVA. 


First, we conducted a brief introduction to the layout and the proposed visualization methods. This stage took approximately thirty minutes for each of the evaluations with the domain experts, after which they were able to use the tool without any further assistance. While both of the domain experts highlighted the intuitiveness of the Hexagon and 3D views, they also pointed out the need for an appropriate documentation describing other views, such as our novel Difference view, in order to support usability. 
According to the domain experts, ChemVA allows to easily compare numerous properties, which is an uncommon but very useful feature in tools for virtual screening. Both of the domain experts highlighted the usefulness of having different projections, and the ability to focus on a specific source of data using the correlation scores and comparing them by means of the Difference view. 
According to the visualization expert, the Difference view does a good job by enabling the user to visualize how the two mappings, hence the two feature spaces, are correlated in terms of grouping the same set of molecules together.
 Both domain experts enumerated several examples from their own ongoing research activities which they claimed could be enhanced by ChemVA, in terms of time and effort For instance, a drug design task aimed at identifying \textit{on- and off-target} drug responses, or the analysis of 4300 FDA-approved drugs for usability in the treatment of COVID-19, which is currently carried out
using cumbersome scripting and manual feature selection. 
 
}

\section{Results and Discussion}
\label{sec:results_and_discussion} 
\rev{In this section we summarize some final remarks and feedback provided by the domain experts who conducted both stages of the evaluation of ChemVA. We discuss the results of both case studies, as well as strengths and limitations of our tool.}
As shown in Section \ref{sec:first_stage}, the goals for both case studies were effectively \rev{fulfilled} using ChemVA, 
\rev{which suggests that the tool has the potential to be widely adopted and used by medicinal chemists. This claim is also supported by the results of the qualitative evaluation performed by external experts, described in Section \ref{sec:second_stage}.}
Our tool \rev{allowed an intuitive comparison of} sets of chemically similar compounds.
In contrast to other tools, which only allow to sort and filter compounds by a set of properties, ChemVA enabled a comprehensive analysis of \rev{the} data by taking into consideration multiple molecular representations and levels of granularity.
\rev{In the case studies, t}his was aided by specific features of \rev{our} tool, such as grouping compounds in hexagons and allowing to adjust their size.

In the first case study, the Detail view assisted the domain expert in distinguishing similar compounds based on proximity in all four projections (R1 \& R2).
Furthermore, the colors and shapes presented in the Hexagonal and Detail views allowed the domain expert to find a specific hexagon very quickly (R1).

The task of analyzing compound similarity could be thoroughly tested by means of different projections provided by ChemVA (R2).
Furthermore, \rev{during the two case studies it was possible} to examine these projections in terms of their trustworthiness, using the provided correlation scores (R3), and to compare them by means of the Difference view (R2). These two features provided by ChemVA aided the domain expert in the first case study in order to assess whether the projections were trustworthy and thus hexagons \rev{were effectively grouping} similar compounds.

The evaluation also showed the usefulness of the 3D view to discover and analyze similar 3D structures among the selected compounds (\rev{R5}).
Examining the chemical determinants for bioactivity in both case studies was straightforward by means of \rev{the} molecular alignment function provided by ChemVA, according to the domain expert.
Every subset of molecules could be quickly aligned on demand, in contrast to other chemical software tools, where this task constitutes a strenuous process and it is often restricted to molecules above a certain threshold of structural similarity \cite{Gutlein2014}.
Moreover, structural patterns could be further explored using the \textit{opacity filter} and \textit{invert opacity option}, which allow highlighting the different parts of the compounds.

\rev{The coordination of the plots with the Table view was also found useful by the domain experts} (\rev{R4}), offering fast access to summary information in its column headers.
The Table view proved to be \rev{effective} for exploring a large dataset, such as \textit{P-glycoprotein}, enabling the search for rare feature values through filtering and hierarchical grouping \rev{during the second case study}.
Other functionalities, such as showing SMILES formulas for copying, the option for exporting 3D structures, or searching compounds by SMILES formula, made ChemVA a flexible tool, easily interoperable with other cheminformatics tools.

One limitation identified \rev{during} the \rev{first} case stud\rev{y} is that new compounds were not always projected near to structurally similar compounds. 
This functionality performed consistently better on some molecular representations than the others, as \rev{it} can be \rev{seen} in Figure~\ref{fig:new_compounds}. 
This might be related to the fact that the dataset serotonin-dopamine used in the experiment is small (118 compounds), thus the neural-based parametric model trained for projecting new compounds is not able to generalize well for every molecular representation. \rev{Another limitation we identified is that not all compounds have available 3D structures, which may limit the usage of the 3D view with some datasets.}

\rev{ChemVA proved to be useful both as an exploration and visualization tool, and it also helped the domain expert with the design and evaluation of new compounds (R6) during the first case study. As discussed in Section \ref{sec:second_stage}, ChemVA was also praised as a useful and innovative tool by the domain experts during the external evaluation.}

\rev{We also identified potential directions for future extensions of the tool, accounting both for user experience and functionality. The external domain experts suggested expanding the current functionality of the tool by enabling the user to load and customize their own data. One of the domain experts also suggested to enable adding new compounds by drawing them in a sketcher. Other possible directions include providing options for substructure search or to enhance navigation by providing breadcrumbs and saving analytical snapshots.}
\rev{Lastly, it is worth noticing that although our tool is domain-specific, ChemVA has been developed by following techniques and strategies that could be applied to other types of data, including non-chemical data. For instance, our newly proposed Difference view could assist in the process of visually analyzing the preservation of clusters and neighborhoods for most clustering methods. Also, the analysis of trustworthiness of DR projections could be applicable to any type of data undergoing a DR procedure. The idea behind the 3D view, which allows for overlapping common molecular substructures, could be applied to other graph-based data or even text.}


\section{Conclusions}
\label{sec:conclusions}
In this paper, we presented ChemVA, a novel tool for visual analysis of chemical compounds, which is especially focused on evaluating molecular similarity for virtual screening. 
Our tool proposes a set of \rev{coordinated} views that support visual exploration and comparison of 2D projections, which are obtained from applying dimensionality reduction on different---and ideally, complementary---molecular representations.
Our case studies, conducted by a medicinal chemist, confirmed that ChemVA addresses the functional requirements and provided appropriate support for his analysis on two different datasets. \rev{The qualitative evaluation conducted by three other external experts revealed the potential of our tool for its adoption in the domain, as well as the usefulness of our newly proposed Difference view. The evaluation process allowed us to identify potential extensions to ChemVA, which will steer our future work on the tool.}


\rev{\section*{Acknowledgments}
\label{sec:acknowledgements}{The presented work has been supported by DFG-GACR research project no. GC18-18647J, by CONICET research grant PIP 112-2017-0100829, by UNS research grants PGI 24/N042 and PGI 24/N048, and by ANPCyT (Argentina) research grant PICT-2017-1246.
We thank Sérgio M. Marques and Ondřej Vávra from the Loschmidt Laboratories, and Michaël Aupetit from the Qatar Computing Research Institute QCRI for evaluating ChemVA and for their insightful comments and suggestions on the features of the tool.}}


\newpage
\section*{\LARGE{Supplementary Material}}
In this Supplementary Material, we provide the readers with detailed information about data preprocessing performed for ChemVA \rev{and computational times,} as well as the results of the molecular docking performed for the Case study. 
Although this document contains important information for 
reproducing our approach\rev{ and the experiments conducted during the evaluation of the tool}, its contents are not crucial for understanding the visualization concepts used in ChemVA, which constitute the core of our proposal.

\section*{Data Preprocessing}\label{sec:data_preprocessing}
\rev{In this section we describe the computation of the vector-based molecular representations and the equations used for computing the correlation scores. We also provide details on the parameterization of our DR method, parametric models and molecular alignment back-end service. The information presented in this section is relevant for reproducibility purposes.}

\subsection*{Vector-based Molecular Representations}
In order to provide domain experts with diverse information about the ligands in our datasets, we computed four different molecular representations, which we then used as sources of data for our 2-dimensional projections. For each dataset, we computed radius 2 ECFPs and Daylight fingerprints, both 1024 bits long, using the \textit{Chem} and \textit{AllChem} packages of RDKit \cite{landrum2016rdkit}. We also computed 0D, 1D, and 2D molecular descriptors using Mordred \cite{moriwaki2018mordred}, obtaining a total of 1613 descriptors. We removed all descriptors that had more than $10\% NaN$ values and replaced all remaining $NaN$ with a fixed value (maximum value per descriptor), which yielded 1454 molecular descriptors. Finally, we used a pretrained Mol2Vec model \cite{jaeger2018mol2vec}, which computed 300-dimensional embeddings based on the SMILES formulas of the compounds in the datasets. 
\rev{The computation of ECFPs, Daylight fingerprints and molecular descriptors took less than an hour, and it was performed in a 32-core cluster with 12GB memory. This preprocessing stage was performed offline, which means that it was done once and then the results were stored for future use.}

\subsection*{Features for Drug-Likeness}

The features related to drug likeness are displayed in the Table view and can be color encoded onto the Detail and Hexagonal views as well. The values for such features were obtained using a publicly available REST API by ChEMBL \cite{gaulton2017chembl}. The collected values did not undergo any further preprocessing steps.

\rev{\subsection*{t-SNE Projections and Parametric Models}

We computed a t-SNE projection for each of the vector-based molecular representations, for both of the datasets used in the case study. The t-SNE projections were computed using the \textit{manifold} class from Scikit-Learn \cite{scikit-learn}, and the parameters used were selected by grid search. The tested perplexity values varied from 5 to 10\% of the total amount of compounds in the dataset. The same parameterization on each dataset was used to fit t-SNE for each of the vector-based molecular representations. Fitting each t-SNE projection took around 20 minutes on a 32-core cluster with 12GB memory. A summary of the parameterization used for t-SNE is provided in Table \ref{tab:params_tsne}.}

\begin{table}[th]
  \caption{\rev{Summary of the parameters used to fit the t-SNE projections of each dataset.}}
  \label{tab:params_tsne}
  \scriptsize%
	\centering%
  \begin{tabu}{ccc}
  \toprule
    Dataset                            & P-glycoprotein & Serotonin-Dopamine \\
  \midrule
    Perplexity                         & 45             & 5$\sim$10         \\
    Maximum No. epochs          & 10000          & 10000              \\
    No. epochs without progress & 500$\sim$1000 & 1000               \\
  \bottomrule
  \end{tabu}
\end{table}

\rev{After fitting each t-SNE projection, we trained one parametric model to learn each of such projections. The parametric models were built using Keras and Tensorflow \cite{chollet2015keras}. The parameterization of these models is summarized in Tables \ref{tab:params_models_g} and \ref{tab:params_models_2t}. The computational time invested on training these models was of approximately 30 minutes per model, on a 32-core cluster with 12GB memory. Both the fitting of the t-SNE projections and the training of the parametric models consisted of offline tasks.}

\begin{table*}[t]
  \caption{\rev{Summary of the parameters used to build and train the four parametric models for \textit{P-glycoprotein} dataset.}}
  \label{tab:params_models_g}
  \scriptsize%
	\centering%
  \begin{tabu}{lcccc}
  \toprule
   \rowfont{\color{black}} Parameters                & ECFPs             & Daylight fps      & Molecular descriptors & Molecular embeddings \\
  \midrule
  \rowfont{\color{black}} \# Nodes per hidden layer & 100 / 10 / 2      & 100 / 10 / 2      & 200 / 20 / 2          & 100 / 10 / 2         \\
\rowfont{\color{black}}Activation fn             & relu              & relu              & relu                  & relu              \\
\rowfont{\color{black}}Dropout coefficients      & 0.25 / 0.15 / 0.1 & 0.25 / 0.15 / 0.1 & 0.25 / 0.15 / 0.1     & 0.25 / 0.15 / 0.1    \\
\rowfont{\color{black}}Early stopping patience   & 70                & 130               & 120                   & 50                   \\
\rowfont{\color{black}}Early stopping delta      & 0.005             & 0.005             & 0.005                 & 0.005                \\
\rowfont{\color{black}}Learning rate (Adam opt)  & 0.0001            & 0.0001            & 0.0001                & 0.0001 \\          
  \bottomrule
  \end{tabu}
\end{table*}

\begin{table*}[t]
  \caption{\rev{Summary of the parameters used to build and train the four parametric models for \textit{serotonin-dopamine} dataset.}}
  \label{tab:params_models_2t}
  \scriptsize%
	\centering%
  \begin{tabu}{lcccc}
  \toprule
  \rowfont{\color{black}} Parameters                & ECFPs             & Daylight fps      & Molecular descriptors & Molecular embeddings \\
  \midrule
   \rowfont{\color{black}} \# Nodes per hidden layer & 50 / 10 / 2      & 50 / 10 / 2      & 200 / 50 / 2          & 50 / 10 / 2         \\
 \rowfont{\color{black}}Activation fn             & relu              & relu              & tanh                  & sigmoid              \\
 \rowfont{\color{black}}Dropout coefficients      & 0.25 / 0.15 / 0.1 & 0.25 / 0.15 / 0.1 & 0.25 / 0.15 / 0.1     & 0.25 / 0.15 / 0.1    \\
 \rowfont{\color{black}}Early stopping patience   & 70                & 70               & 100                   & 70                   \\
 \rowfont{\color{black}}Early stopping delta      & 0.005             & 0.005             & 0.005                 & 0.005                \\
 \rowfont{\color{black}}Learning rate (Adam opt)  & 0.0001            & 0.0001            & 0.0001                & 0.0001 \\          
  \bottomrule
  \end{tabu}
\end{table*}

\subsection*{Correlation Scores for Measuring the Trustworthiness of Projections}
t-SNE applies two different distributions to map data points to a lower-dimensional space in an effort to preserve neighborhoods according to a specific optimization criterion. As a result, pairwise distances among the compounds in the low-dimensional space could be distorted with respect to distances in the high-dimensional space, and those differences could mislead the domain expert during the virtual screening process. For this reason, as a means to assess the trustworthiness of the projections, we computed two different \textit{correlation scores} for each of the compounds. 

The first score, namely $r$, was computed by calculating the cosine distance of each compound $k$ to the rest of the compounds in the high-dimensional space and then measuring the Pearson correlation (see Equation \ref{eq:pearson}) among these distances and the ones in the low-dimensional space. Pearson correlation has been previously used by Strickert et al in MSR~\cite{strickert2010adaptive}, which has also been applied to chemical compounds. The second score, namely $\tau$, was computed by generating two ranks for each compound $k$, one in the origin space and another one in the mapped space, which sort all the compounds according to the cosine distance to $k$. Afterwards, the score was computed by comparing both ranks using the Kendall rank correlation (see Equation \ref{eq:kendall}).

Both correlation scores were computed for each compound individually, and once for each of the four vector-based molecular representations used in ChemVA, which yields four different views on which the trustworthiness of each projection can be assessed. In addition, the \textit{difference graph} uses the Kendall rank correlation scores as the default encoding. These scores were computed using tools in the Python Data Science suite SciPy \cite{scipy} and Scikit-Learn \cite{scikit-learn}.

\begin{equation}
\label{eq:pearson}
r = \frac{n\left ( \sum xy \right )-\left ( \sum x \right )\left ( \sum y \right )}{\sqrt{\left [n\sum x^{2} - \left ( \sum x \right )^{2} \right ]\left [n\sum y^{2} - \left ( \sum y \right )^{2} \right ]}}
\end{equation}

\begin{equation}
\label{eq:kendall}
\tau = \frac{n_{concordants}-n_{discordants}}{n\left ( n-1 \right )/2}
\end{equation}

\subsection*{Molecular Alignment and 3D Conformations}
ChemVA supports a 3D view, which allows the user to conduct a detailed analysis of the molecular structures under study, and to visually explore common geometrical patterns among compounds. The 3D conformations of compounds were retrieved from PubChem \cite{kim2016pubchem} and computed using OpenBabel \cite{OBoyle2011}. The computed 3D conformations were obtained using \textit{Ghemical} force field energy minimization, force field cleanup (500 cycles) and slow rotor search.
\rev{The computation of these 3D conformations consisted of an offline task. It took approximately between 2 and 20 minutes for each compound, depending on the complexity of its molecular structure, and it was performed on a 32-core cluster with 12GB memory.}

The molecular alignment function involves two steps. The first step consists of finding the \textit{maximum common substructure} (MCS) among the selected compounds, which is achieved using RDKit \cite{landrum2016rdkit}. The second step consists of aligning all the selected molecules with respect to the found MCS, and it is done using the \textit{obfit} function provided by OpenBabel \cite{OBoyle2011}. \rev{The molecular alignment task is performed \textit{on demand}: it is an online task, which means that it is performed upon the selection of a set of compounds in the Detail view in our back-end servers. The response time of this web service is of approximately one second, tested on random selections from 5 to 30 compounds.}

\rev{
\subsection*{Adding New Compounds}
The addition of new compounds to the Detail view is an online task, performed \textit{on demand} in our back-end servers. The parametric models are previously loaded, which reduces significantly the response time: approximately four seconds per compound.}

\section*{Case study 1 - Molecular Docking}

This \rev{s}ection provides details \rev{on the two datasets used in the case studies, as well as} on the results of the molecular docking study performed by the domain expert in the context of our case study 1 (see  Section 6 of the paper). 

\rev{\subsection*{Dataset Selection and Preprocessing}

The two datasets used in the case studies, namely \textit{serotonin-dopamine}\footnote{https://www.ebi.ac.uk/chembl/target\_report\_card/CHEMBL214/}\footnote{https://www.ebi.ac.uk/chembl/target\_report\_card/CHEMBL217/} and \textit{P-glycoprotein}\footnote{https://www.ebi.ac.uk/chembl/target\_report\_card/CHEMBL4302/}, were built upon two different datasets retrieved from ChEMBL \cite{gaulton2017chembl}.
We performed a preliminary selection step of the SMILES formulas on each of the datasets, keeping only those compounds that could be loaded using RDKit \cite{landrum2016rdkit}. Details on the amount of compounds belonging to each class are provided in Tables \ref{tab:details_g} and \ref{tab:details_2t}. The compounds were labeled according to their experimentally measured \textit{IC50} bioactivity value towards the targets under study. 
}

\begin{table}[ht]
  \caption{\rev{Details on the composition per class of the \textit{P-glycoprotein} dataset.}}
  \label{tab:details_g}
  \scriptsize%
	\centering%
\begin{tabular}{>{\color{black}}c>{\color{black}}c>{\color{black}}c}
\hline
\multicolumn{1}{l}{}                                                               & Class & P-glycoprotein \\ \cline{2-3} 
\multirow{4}{*}{\begin{tabular}[c]{@{}c@{}}\# compounds \\ per class\end{tabular}} & Active      & 42             \\
                                                                                   & Moderately active & 178            \\
                                                                                   & Inactive   & 673            \\ 
                                                                                   & \textbf{Total}                       & \textbf{893}   \\ \hline
\end{tabular}
\end{table}

\begin{table}[ht]
 \caption{\rev{Details on the composition per class of the \textit{serotonin-dopamine} dataset.}}
  \label{tab:details_2t}
  \scriptsize%
	\centering%
\begin{tabular}{>{\color{black}}c>{\color{black}}c>{\color{black}}c>{\color{black}}c}
\hline
\multicolumn{1}{l}{}                                                               & Class      & Serotonin 1a & Dopamine D2  \\ \cline{2-4} 
\multirow{4}{*}{\begin{tabular}[c]{@{}c@{}}\# compounds \\ per class\end{tabular}} & Active       & 14           & 5            \\
                                                                                   & Moderately active  & 42           & 28           \\
                                                                                   & Inactive     & 62           & 85           \\ 
                                                                                   & \textbf{Total}                       & \textbf{118} & \textbf{118} \\ \hline
\end{tabular}
\end{table}

\subsection*{Preparation of Receptor Structures}
The structure of human 5HT1A receptor was constructed using homology modeling using the \textit{I-TASSER} web server using default settings \cite{yang2015tasser}. The model was constructed using homologous 5HT receptors and monoamine receptors with resolved structures. Model 1 was selected for the docking study with a TM-score of $0.52$ ± $0.15$, indicating a well-predicted fold \cite{zhang2004scoring}. The model was aligned to the homologous structure of 5HT1B (PDB ID 4IAR \cite{wang2013structural}) and showed a nearly perfect overlap. The structure of the dopamine D2 receptor was downloaded from PDB \cite{berman2000rcsb} (PDB ID 6CM4\cite{wang2018structure}). Hydrogens, ions, and waters were removed using PyMOL \cite{PyMOL}. The protonation state of titratable groups, as well as the orientation of Asn, Gln, and His groups, were corrected by the H++ web server using pH 7.5 \cite{anandakrishnan2012h}.

\subsection*{Preparation of Ligand Structures}
Based on the three structures selected in ChemVA, five new ligand structures were designed using Avogadro \cite{hanwell2012avogadro}. Molecular docking aimed to compare binding energies and binding modes of the three dataset structures and those of the five designed structures. The molecular docking was done using Autodock Vina \cite{trott2010autodock}. Autodock atom types and Gasteiger charges were added to the two receptors and the ligands using MGLTools \cite{morris2009autodock4, sanner1996reduced}. The docking grid was selected to be a $30$ x $30$ x $30$ $Å$ cube. For the 5HT1A receptor, the cube was centered on the bound ligand in the aligned structure of 5HT1B. For the dopamine D2 receptor, the cube was centered on the bound ligand in the receptor structure.

\subsection*{Results and Discussion}
All of the seven structures successfully bound to the binding pockets of the two receptors, suggesting a plausible binding mode was found. Three of the five designed structures showed a lower value binding energy, indicating a stronger binding. The binding energies of the best binding modes are shown in Table \ref{tab:docking}\rev{, and the SMILES formulas for each of the designed compounds are provided in Table \ref{tab:smiles}}.

\begin{table}[th]
\caption{Predicted binding affinities towards the \textit{serotonin 5HT1A receptor (5HT1A)} and \textit{dopamine D2 receptor (dopD2)} of both selected dataset compounds and designed compounds, measured in $kcal/mol$. Rows in bold show the compounds exhibiting the highest predicted affinities to both receptors.}
\label{tab:docking}
\centering
\begin{tabu}{ccc}
\toprule
{Compound}                     & {Affinity to 5HT1A}  & {Affinity to dopD2 } \\ \midrule
{Compound 45}                  & {-8.0}                               & {-8.7}                               \\
{Compound 79}                  & {-8.0}                               & {-8.3}                               \\
{Compound 117}                 & {-9.9}                               & {-11.0}                              \\
{Designed compound 1}          & {-8.4}                               & {-11.5}                              \\
{\textbf{Designed compound 2}} & {\textbf{-10.5}}                     & {\textbf{-11.9}}                     \\
{Designed compound 3}          & {-9.9}                               & {-10.8}                              \\
{\textbf{Designed compound 4}} & {\textbf{-10.3}}                     & {\textbf{-11.2}}                     \\
{\textbf{Designed compound 5}} & {\textbf{-10.2}}                     & {\textbf{-11.2}}                     \\ \bottomrule
\end{tabu}
\end{table}

\begin{table}[th]
\caption{\rev{SMILES formulas of the designed compounds. Rows in bold show the compounds exhibiting the highest predicted affinities to both receptors.}}
\label{tab:smiles}
\centering
\resizebox{\columnwidth}{!}{%
\begin{tabu}{>{\color{black}}c>{\color{black}}c}
\toprule
{Designed Compound}                     & {SMILES formula} \\ \midrule
{1}          & {Fc1ccc(cc1)c2cncc(CNCC3CCc4ccccc4C3)c2}           \\
{\textbf{2}} & {\textbf{Fc1ccc(cc1)c2cncc(CNCC3CCc4cc(C)c(C)cc4C3)c2}} \\
{3}          & {N(Cc1cncc(c1)c2cc(F)ccc2F)CC3CCC4=C(C=CC=C4)O3}           \\
{\textbf{4}} & {\textbf{N(Cc1cncc(c1)c2ccc(F)cc2)C(N(C)C)C3CCC4=C(C=CC=C4)O3}} \\
{\textbf{5}} & {\textbf{Fc1ccc(cc1)c2cncc(c2)C(N)CCC3CCc4ccccc4O3}} \\ \bottomrule
\end{tabu}}
\end{table}

In this comparison, designed compounds \textbf{2}, \textbf{4}, and \textbf{5} exhibited higher predicted affinities to both receptors. These predicted results do not necessarily mean higher inhibitory activity, given that such bioactivity can be influenced by other factors such as flexibility of the receptor proteins. 


In addition, we provide supplementary files comprising the results from the I-TASSER web server and the input files and results from the Autodock Vina procedure.

\vspace{10mm}
\bibliographystyle{abbrv-doi}
\bibliography{references}
\end{document}